\documentclass[11pt,a4paper]{article}
\pdfoutput=1
\usepackage{jinstpub}
\usepackage[sort&compress]{natbib}
\usepackage{graphicx}
\usepackage{epsfig}\usepackage{dcolumn}% Align table columns on decimal point
\usepackage{algorithm}
\usepackage[noend]{algpseudocode}

\usepackage[utf8]{inputenc}
\usepackage[T1]{fontenc}
\newcommand{\met} {{E\!\!\!\!/_{\rm T}}}

\newcommand{ \sklearn } {{\tt scikit-learn }}
\newcommand{ \xgboost } {{\tt XGBoost }}
\newcommand{ \keras } {{\tt Keras }}
\newcommand{ \tmva } {{\tt TMVA }}

\makeatletter
\def\BState{\State\hskip-\ALG@thistlm}
\makeatother

\title{\bf Stacking machine learning classifiers to identify Higgs bosons at the LHC}

\author[a]{Alexandre Alves}
\affiliation[a]{Departamento de Física, Universidade Federal de São Paulo, Diadema-SP, 09972-270, Brazil}
\emailAdd{aalves@unifesp.br}
\abstract{
Machine learning (ML) algorithms have been employed in the problem of classifying signal and background events with high accuracy in particle physics. In this paper, we compare the performance of a widespread ML technique, namely, \emph{stacked generalization}, against the results of two state-of-art algorithms: (1) a deep neural network (DNN) in the task of discovering a new neutral Higgs boson and (2) a scalable machine learning system for tree boosting, in the Standard Model Higgs to tau leptons channel, both at the 8 TeV LHC. In a cut-and-count analysis, \emph{stacking} three algorithms performed around 16\% worse than DNN but demanding far less computation efforts, however, the same \emph{stacking} outperforms boosted decision trees. Using the stacked classifiers in a multivariate statistical analysis (MVA), on the other hand, significantly enhances the statistical significance compared to cut-and-count in both Higgs processes, suggesting that combining an ensemble of simpler and faster ML algorithms with MVA tools is a better approach than building a complex state-of-art algorithm for cut-and-count.}


\arxivnumber{1612.07725}

\begin{document}
\maketitle
\flushbottom 

%%%%%%%%%%%%%%%%%%%%%%%%

\section{Introduction} 
%Artificial Intelligence (AI) algorithms are becoming essential in many areas of contemporary life. Training them to make accurate predictions beyond human capabilities is desirable in several dedicated tasks which, in principle, cannot be accomplished by humans. This is the case of discovering new particles in high energy colliders. Categorizing a fireworks pattern showing up in a particle detector as either a new particle or just another  Standard Model (SM) manifestation is well beyond what we are able to do.  

Artificial Intelligence (AI) algorithms are becoming important in many areas of contemporary science. This is the case of discovering new particles in high energy colliders. Much of the work of the particle physicists in predicting which collision events are interesting signals and which are backgrounds is still based on visual identification of signal rich regions comparing histograms of kinematic distributions.
%, the well known and widespread \emph{cut-and-count} technique.
On the other hand, searching for particles signals in colliders is a typical binary classification problem which can be tackled by AI methods. Actually, machine learning algorithms have been employed for a long time to classify signal and background events in a more efficient way compared to cut-and-count methods. Boosted decision trees (BDT), Neural Networks (NN) and naive Bayes (NB) are the most used classifiers in particle physics but, state-of-art algorithms as \emph{deep neural networks} (DNN)~\cite{dnn} and \emph{scalable machine learning system for tree boosting}~\cite{xgboost} have been proposed recently. 

\emph{Baldi et al} demonstrated that DNN works extremely well in the search for SM and beyond SM (BSM) Higgs bosons at the 8 TeV LHC~\cite{Baldi:2014kfa, Baldi:2014pta}. Moreover, they showed that a DNN is able to learn high-level representations of the data almost matching the human expertise in designing discriminating distributions. Other applications of the DNN were studied in jet tagging~\cite{Baldi:2016fql, Guest:2016iqz, Kasieczka:2017nvn}, neutrino physics~\cite{Aurisano:2016jvx}, galaxy classification~\cite{Kim:2016knv}, and supernovae classification~\cite{Charnock:2016ifh}, for example. 

In spite of their power, DNNs are hard to be trained and very computationally demanding. Tuning the hyperparameters of this class of algorithms in order to get an optimal performance requires parallel computations either with GPUs or CPU clusters due their high complexity. Another important ingredient which seems to be necessary for high classification accuracy is having a large training set possibly with millions of simulated events. The simulation of collision events, by its turn, might be very time consuming.

However, much before the rise of deep neural networks, data scientists knew how to improve the performance of single classifiers by combining them in ensembles and committees~\cite{Zhou} with important advantages compared to single weaker leaners. For example, training an algorithm on a given set of examples does not guarantee good classification performance on new data, if the algorithm is too complex it might learn all the small idiosyncrasies of the training set but commit gross errors on unseen data with different fluctuations, in this case the \emph{generalization error} is large, the algorithm overfits the data. Keeping overfitting under control is one of the major goals of ensemble methods like \emph{Boosting}~\cite{boosting} and \emph{Bagging}~\cite{bagging}. Choosing the best classifier for a given problem is not easy too, a good performance classifier trained on a given training data set might perform not so good on all possible new data sets, combining many learners helps to make the final classifier more robust and less sensitive to fine tunings~\cite{Zhou2009}. \emph{Boosting}, by the way, is crucial for decision trees to be useful and has been implemented in various public ML packages as \tmva~\cite{tmva} whose boosted decision trees (BDT) algorithm has been widely used in experimental studies of ATLAS and CMS. 
%It is interesting to note that many ML contests are won by solutions based on ensembles. 
%For example, the winning solution of the famous Netflix preference prediction contest~\cite{netflix} was a very complex ensemble, so complex that the company decided not implement it for commercial purposes. In scientific applications though we are not faced with such practical limitations.

A possible issue concerning ML is that a given learner probes only part of the whole hypothesis space. The basic learning problem is to find the true function that predicts outcomes given the information available, the space of all possible functions that might be a solution to this problem is the hypothesis space. Picking a single classifier implies choosing a class of functions which might have, in principle, nothing to do with the true function of the problem. In this respect, combining heterogeneous classifiers provides good base functions to approximate the true function~\cite{Zhou2009}.

From the statistical point of view, ignoring the correlations among the kinematic distributions decreases the power of the statistical tests. This was nicely illustrated in the case of discriminating the Higgs boson spin hypothesis in the $ZZ$ channel~\cite{DeRujula:2010ys}. It is not guaranteed that a ML algorithm is able to learn all those correlations and represents them in an one dimensional distribution, thus training more than one classifier might be a better approach to capture the correlations and increase the power of the statistics tests based on ML outputs. On other hand, reducing a high-dimensional distributions space to a low-dimensional one, with possibly better discriminating distributions, should make it easier the exact computation of the likelihood ratios by sampling this new low-dimensional probability distribution function (PDF).

In this work, we show that an ensemble learning technique called \emph{stacked generalization}~\cite{Wolpert}, \emph{stacking} for short, is able to produce a strong learner from weaker ones with rather modest tuning in the search for two distinct Higgs boson production processes previously studied in the literature~\cite{Baldi:2014kfa,kaggle}. However, instead of just comparing algorithms, we aim to compare entire approaches to the discovery prospects of those particles and interactions. In this respect, we will find that combining the stacked classification outputs of the events with multivariate statistical analysis outperforms the state-of-art algorithms used for cut-and-count. 
%compares to the state-of-art ML algorithms employed in the search for two distinct Higgs boson production processes previously studied in the literature. 

First, for the gluon fusion production of a new heavy neutral Higgs boson decaying to a charged scalar and a $W$ boson  with the Run I data of the 8 TeV LHC, we will show that stacking demands much less computational resources and running time required for training, validation and optmizations than those necessary to tune and train a DNN, at the cost of a mild decrease in statistical significance computed by cutting on the scores distribution to clean up the backgrounds as done in Ref.~\cite{Baldi:2014kfa}. Switching to a multivariate analysis should, in principle, give better results. For this later goal, constructing more than one ML output to classify the events is likely to better capture correlations which enhance the statistical significances.  By substituting the high-dimensional kinematic distributions space by a low-dimensional ML outputs space, MVA performs better, making it possible to sample the multidimensional PDF of the events much more easily to take advantage of the benefits of exploring the correlations often neglected in MVA. This last approach greatly enhances the signal significance compared to those of Refs.~\cite{Baldi:2014kfa} as we are going to show.

Second, we compare the results of stacking against those of the 2014 Higgs Machine Learning Challenge, a contest hosted by Kaggle~\cite{webkaggle} in the search for the SM Higgs to tau leptons in the gluon fusion channel at the 8 TeV LHC~\cite{kaggle} proposed by the ATLAS Collaboration. In special, we aim comparisons against the fast high performance decision trees algorithm implementation \texttt{XGBoost}~\cite{xgboost}, winner of the \emph{HEP meets ML} award of the contest. In this case, we found that stacking ML outputs over the top of \texttt{XGBoost} performs always better, both in the cut-and-count analysis and MVA, with similar computation time and less care with the tuning of the algorithms. 
%By the way, the winner solution of the contest was an ensemble of 70 DNNs~\cite{gabor}.

We have to point out that stacking classifiers is not an entirely new idea in particle physics. The CDF Collaboration used an ensemble of ML tools to discover the single top production process at the Tevatron~\cite{Aaltonen:2010jr}. Other applications of ensemble of ML classifiers can be found, for example, in the primary trigger system~\cite{Likhomanenko:2015aba} and in the particle identification algorithm of the LHCb~\cite{PID}. Also, in jet tagging, the final classifier for a given type of particle is built upon several pieces of information, part of them is the output of ML classifiers which are employed to improve the performance of that final classifier.

In this work, we guide ourselves by the literature of the subject, confirming some important findings as we are going to discuss in the next sections. Ensemble techniques were also used extensively in 2014 HiggsML contest as discussed in Ref.~\cite{kaggle}.

Both the MVA performed directly on the ML outputs of this experimental work as the analysis of our work are justifiable in view of Ref.~\cite{Cranmer:2015bka}. One possible issue of stacking many ML outputs is the estimation of systematic uncertainties which should be propagated to the calculation of the statistical significance of the signals, anyway Refs.~\cite{Aaltonen:2010jr,Likhomanenko:2015aba} show that this is feasible by the experimental collaborations.

From now on, we refer ourselves to the $H\to H^\pm W^\mp$ process with DNNs as the BSM Higgs process, and to the $h\to\tau^+\tau^-$ process with \texttt{XGBoost} as the SM Higgs process.

The paper is organized as follows: in section II we describe stacking, the collisions process which are targets of the classification tasks, how we train and test the algorithms and present our first results; section III is devoted to experiments with shallower deep neural networks using the stacked outputs; in section IV we present a dedicated multivariate analysis; finally, in section V we present our conclusions.

\section{Stacking machine learning classifiers}
%\subsection{Description of the algorithm}

\subsection{Description of the algorithm} There are many ensemble approaches in machine learning, see~\cite{Zhou}. Perhaps, the most well known are boosting and bagging algorithms used to reduce overfitting of decision trees. Amongst the most simple ideas is casting votes and deciding a class label by the vote of the majority. 
%Suppose we train some algorithms for a common classification problem, none of them too accurate in its task. Casting the votes of all classifiers and basing the decision on the majority usually helps to decrease the classification error. 
Another simple algorithm is just averaging the estimate of the posterior probabilities of each class label classification.
%~\cite{Zhou}.

Another possible approach is to train a number of classifiers and use their outputs as features to train a new classifier called a \emph{generalizer}~\cite{Wolpert}. In this respect, the classifiers trained on the original kinematic distributions actually work as dimensionality reduction maps. Good performance maps make explicit the differences of the original distributions in the new one-dimensional binned distribution, the classifier output. However, it might be that different classifiers are better at perceiving correlations between certain kinematic distributions than others, that is why combining various maps is likely to complement each others performance as discussed in the Introduction. 
These types of ensemble can also be understood as a high-level feature engineering. A prime concern in machine learning is how to represent data to be used as training examples for an algorithm. Raw data can always be preprocessed in order to exhibit hidden correlations. In particle physics, the distributions to be chosen to represent a given data set are constructed based on physical intuition and reasoning and by experience acquired on previous cases. Some of them are absolutely essential for particle discovery, as invariant masses, others are less transparent but useful. Representing data as histograms of classifiers outputs is nothing but complex and sometimes physically opaque new distributions built upon other possibly more physically intuitive ones. What really matters is that these new distributions often present a much superior discerning power compared to the original physically transparent distributions.

%The algorithm that we propose here is that used in~\cite{Aaltonen:2010jr} which, by its turn, is known as \emph{stacking}~\cite{Wolpert}. We could also called it a \emph{mixture of experts} and other nominations in the ML field exists. 
%As our procedure is a generalization of stacking we prefer to keep this name even because the original idea of Ref.~\cite{Wolpert} is just a particular case of this general approach. 
The pseudo-code that summarizes the stacking of learners is shown in Algorithm~(\ref{figstack})~\cite{Wolpert, Zhou2009}.

%\begin{algorithm}[H]
%\begin{algorithmic}
%\State Hi
%\end{algorithmic}
%\end{algorithm}
%The original stacking algorithm of Ref.~\cite{Wolpert} corresponds to the three first {\bf procedures} of this 3-level Stacking recipe of Fig.~(\ref{fig:stack}) with just two level-0 data sets ${\cal D}_{00}$ and ${\cal D}_{10}$, an ensemble $N$ of level-0 classifiers and just one generalizer whose outputs would correspond to ${\cal D}_{11}$. 
%
\begin{algorithm}[H]
\caption{Stacking}

\begin{algorithmic}[t]
% \State 1
\State {\bf Input from simulations:}
\State \emph{Create $3$ level-0 data sets with kinematic features $\mathbf{x}$} 
\State ${\cal D}_{i0}=\{(\mathbf{x}^{(i)}_I,y^{(i)}_I),I=1,\cdots,n\},\; i=0,1,2$

\Procedure{Train level-0 classifiers}{}
\State \emph{Choose a level-0 ensemble of algorithms: ${\cal L}^{(0)}_k$} 
\State \emph{Train level-0 ensemble with training data: ${\cal D}_{00}$}
\State $h_k^{(0)}={\cal L}^{(0)}_k({\cal D}_{00}),\; k=1,\cdots, N$
\State $C_0(\cdot)=\left(h^{(0)}_1(\cdot),\cdots,h^{(0)}_N(\cdot)\right)$
\EndProcedure

\Procedure{Classify level-0 test sets}{}
\State $C_0({\cal D}_{i0})=\left\{\left(C_0(\mathbf{x}_I^{(i)}),y_I^{(i)}\right),I=1,\cdots,n\right\},\; i=1,2$
\EndProcedure

\State {\bf Intermediate Output:}
\State \emph{Create level-1 sets with level-0 outputs as features}
\State ${\cal D}_{i1}=C_0({\cal D}_{i0}), i=1,2$

\Procedure{Train level-1 classifier}{}
\State \emph{Choose generalizer: ${\cal L}^{(1)}_1$ }
\State \emph{Train generalizer with the level-1 training data: ${\cal D}_{11}$}
\State $h_1^{(1)}={\cal L}^{(1)}_1({\cal D}_{11}),\; k=1,\cdots, N$
\State $C_1(\cdot)=\left(h^{(1)}_1(\cdot)\right)$
\EndProcedure

\Procedure{Classify level-1 test set}{} 
\State $C_1({\cal D}_{21})$
\EndProcedure

\State{\bf Final Output:}
\State \emph{Write the final output}
\State ${\cal D}_{22}=\left\{\left(C_1(C_0(\mathbf{x}_I^{(2)})),y_I^{(2)}\right),I=1,\cdots,n\right\}$ 

\end{algorithmic}
\label{figstack}
\end{algorithm}
To give a concrete example, in the experimental single-top search of Ref.~\cite{Aaltonen:2010jr}, the CDF Collaboration trained three ML level-0 algorithms: a boosted decision tree, a single layer neural network and a naive Bayes classifier, and also assembled the output of the matrix-element method to this level-0 classifiers. It was reported that each of those methods performs better than cut-and-count but none of them is able to reach the statistical level for discovery. Then comes stacking. The outputs of those four algorithms were then used to train a generalizer, a single layer neural network, which performed better than the single best classifier of the four level-0 classifiers and reached $5\sigma$ of significance. 

After all that processing of the original kinematic distributions, the final classifier associates to each physical collision event an \emph{output score} $h(\mathbf{x})\in [0,1]$ which is an estimate of the probability of being signal or background. Histogramming the events according to the probability of being signal, for example, calibrates efficiently these final distributions which are used for all the statistical analysis including the marginalization over all the systematic uncertainties of the experiment~\cite{Cranmer:2015bka}.

%The algorithm proposed in Fig.~(\ref{fig:stack}) is essentially the same adopted by the CDF Collaboration in the search for single top quarks. 
The final set of outputs after stacking, ${\cal D}_{22}$ in Algorithm~(\ref{figstack}), is then meant to substitute or augment the original kinematic distributions $\mathbf{x}$ in the estimation of the binned log-likelihood ratio (LLR) statistics in the signal \emph{versus} background hypothesis test or in a cut-and-count analysis. The procedure could be extended to include more level-1 classifiers or even a level-2 classifier trained on the level-1 outputs. Actually, apart from possible issues concerning evaluation of systematic uncertainties, there's no limit in the number of layers of high-level classifiers which can be built. By the way, concerning systematics, an approximation like the one proposed in Ref.~\cite{Fichet:2016gvx} could be used when a large number of small sources of uncertainties is present.
%\begin{eqnarray}
% {\cal D}_{22} &=& \left\{\left((h^{(1)}_1({\cal D}_{21}),y^{(2)}_I\right),I=1,\cdots,n \right\}\nonumber\\
% &=&  \left\{\left(C_1(C_0(\mathbf{x}_I^{(2)})),y_I^{(2)}\right),I=1,\cdots,n\right\}
%\end{eqnarray}

\subsection{Discriminating Higgs bosons events from backgrounds at the 8 TeV LHC} 
%One of the motivations of studying this approach is to compare our results with state-of-art ML algorithms. 
Let us now describe the two Higgs search processes which we are going to study.

\vskip0.5cm
\noindent\underline{(1) BSM Higgs boson}
\vskip0.5cm
  The DNNs of Ref.~\cite{Baldi:2014kfa} were used to discover a heavy neutral Higgs boson in the gluon fusion process
\begin{equation}
pp\to H\to H^\pm W^\mp\to W^\pm W^\mp h\to jjb\bar{b}\ell^\pm+\met
\label{bsmH}
\end{equation}
in the context of the 8 TeV LHC. 

This process is expected in extended Higgs sectors as the Two Higgs Doublets Model~\cite{Djouadi:2005gj, Branco:2011iw}, where five types of Higgs bosons are predicted including a SM-like Higgs $h$, a heavier neutral (pseudo)scalar $(A)H$ and charged scalars $H^\pm$. In the benchmark model assumed in Ref.~\cite{Baldi:2014kfa}, the mass of $H$($H^\pm$) is 425(325) GeV. The main background for this process is the $t\bar{t}$ production with semileptonic decays.

Eleven million events, 53\% of signals and 47\% of backgrounds, were simulated and are available in~\cite{uciBaldi}. The attributes of the events comprise 28 distributions, 21 low-level variables as transverse momentum and rapidity of charged leptons and jets from the tau leptons decays and total missing transverse energy, and 7 high-level variables derived from the low-level ones as invariant mass distributions, angular distance and azimuthal differences between particles among others; more detailed information can be obtained in Ref.~\cite{Baldi:2014kfa}. Given the cuts and detector efficiencies and the Run I luminosity, the authors of Ref.~\cite{Baldi:2014kfa} estimate 100 signal events and 1000 background events with a 5\% systematics in the number of background events.

 We found no information about the significance metrics used in Ref.~\cite{Baldi:2014kfa} but, as the authors quote a systematic uncertainty in the background ($\varepsilon=0.05$), and the number of signal events is much smaller than backgrounds we adopt the following two approximate median significance (AMS) functions for our computations and comparisons in cut-and-count analysis
\begin{equation}
\hbox{AMS}_1=\frac{S}{\sqrt{S+B}}\;\; \hbox{and}\;\; \hbox{AMS}_2=\frac{S}{\sqrt{B+(\varepsilon\times B)^2}}
\label{ams}
\end{equation}
where $S$($B$) is the number of signal(backgrounds) events after cuts. Actually, $\hbox{AMS}_1$ reproduces the result of Ref.~\cite{Baldi:2014kfa} with the signal and backgrounds efficiencies quoted in that work.

%The classification problem which we want to attack is therefore of discriminating $H^\pm W^\mp$ events which come from the decay of a heavy Higgs boson $H$ produced via gluon fusion. 

\vskip0.5cm
\noindent\underline{(2) SM Higgs boson}
\vskip0.5cm

 An approach using boosted decision trees was found to be the most useful algorithm for the ATLAS Collaboration in the 2014 HiggsML Kaggle contest in the search for SM Higgs to tau leptons~\cite{chen}
\begin{equation}
pp\to h\to \tau^+\tau^-\;\; 
\label{smH}
\end{equation}
at the 8 TeV LHC. The main background to this process is the electroweak pair production of tau leptons.

The detection of SM Higgs bosons in the tau leptons channel has been observed with $5.5\sigma$ after ATLAS and CMS combine the data of their 7 and 8 TeV runs~\cite{Khachatryan:2016vau}. Back to 2014, however, the situation was different and there was just an evidence of around $3\sigma$ at that time. Concerned in seeking for better tools to help increasing the significance in that channel, the ATLAS Collaboration proposed a contest at the Kaggle web page in order to learn a better approach from the machine learning community.

The algorithm found most useful for future applications was given by a carefully tuning of the fast and powerful boosted decision trees implementation \xgboost~\cite{xgboost}. This solution was able to reach around $3.72\sigma$ against $3.81\sigma$ of the top performance solution of Ref.~\cite{gabor}, a bagging of 70 DNNs. By the way, the second best solution was found by stacking a large number of BDT outputs~\cite{tim}.

Around 850000 events for Higgs to tau leptons from the ATLAS Collaboration simulation for signal and backgrounds are available at this location~\cite{uciKaggle}. The features comprise 30 low and high-level distributions~\cite{kaggle} for a complete description. The number of signal and background events estimated by ATLAS in the Run I are 682 and 411000 events, respectively for a tiny $1.7\times 10^{-3}$ signal to background ratio. No systematic uncertainties were assumed in the contest. 

The statistical significances were computed from the approximate median significance given by~\cite{kaggle}
\begin{equation}
AMS_3=\sqrt{2}\sqrt{S-(S+B)\ln\left(1+\frac{S}{B}\right)}\; .
\label{ams3}
\end{equation}

The objective function which we optimize in the training of all algorithms is the classification accuracy but our final goal is to increase AMS cutting on the classifier output, and the statistical signal significance calculated from the distribution of log-likelihood ratio statistics based on the ML outputs built with pseudoexperiments.

\subsection{Training and testing the algorithms}
Many works on ensemble of machine learning algorithms suggest that diversifying the algorithms used in a combination of classifiers is a key ingredient to get good results~\cite{Kuncheva}. Also the type of output necessary to train the high-level classifiers have an impact on the final performance. In this respect, the authors of Ref.~\cite{Witten} found that using the estimates of the posterior probabilities of the class labels is better than using the class label predictions themselves (0 or 1). Moreover, they also investigated the performance of various high-level classifiers and found that linear regression models are favored.

Three different algorithms were used as level-0 classifiers: an extreme gradient boosted trees (XGB$_0$) with \xgboost~\cite{xgboost}, a single layer neural network (NN$_0$) with \keras~\cite{keras} with \texttt{Theano}~\cite{theano} back end and a private implementation of a naive Bayes classifier (NB$_0$) as described in Ref.~\cite{Aaltonen:2010jr}. We actually tested other algorithms but those three was found to provide more stable and fast results. We also tested many generalizers, but the best level-1 generalizer in terms of performance and running time, confirming~\cite{Witten}, was a logistic regression classifier (LR$_1$) from \sklearn~\cite{sklearn}. Further details about the models used in each of these classifiers can be found in the vast ML literature, for example, in~\cite{Hastie, Duda, Haykin, Breiman}. 

%We used \xgboost~\cite{xgboost} to obtain the boosted decision trees classification. It is a fast (due easy parallelization) and scalable implementation which performs very well in many ML applications, for example, in the winning solution to the Higgs Kaggle challenge proposed by the ATLAS Collaboration back to 2014~\cite{Chen}. In that challenge, the objective was to obtain the highest AMS for a given output cut for the Higgs to tau leptons at the 8 TeV LHC with simulated events provided by the collaboration. 
%The winning solution AMS reached around 3.7$\sigma$ using just \xgboost and clever distributions engineering.

%For the single layer neural network we used \keras~\cite{keras}, a user friendly, modular package aimed to build deep neural networks on top of \texttt{Theano}~\cite{theano}. 
%or \texttt{TensorFlow}~\cite{tf}. 

%The third level-0 algorithm chosen was a naive Bayes classifier as described in Ref.~\cite{Aaltonen:2010jr}. A naive Bayes classifier, contrary the other two, is a generative algorithm which estimates the class probabilities themselves. We used a private implementation which assumes no specific probability distribution for the features. Further details about the models used in each of these classifiers can be found in the vast ML literature, for example, in~\cite{Hastie, Duda, Haykin, Breiman}.

Let us describe the generalizer briefly. A logistic regression classifier is just a linear model whose output is computed from the weighted combination of the inputs, in this case, the three level-0 ML scores $h(\mathbf{x})$ for a given features vector $\mathbf{x}$
\begin{equation}
f(\mathbf{x}) = w_0+w_1 h_1^{(0)}(\mathbf{x})+w_2h_2^{(0)}(\mathbf{x})+w_3h_3^{(0)}(\mathbf{x})\; .
\end{equation}

The weights $w$ are chosen in order to minimize a regularized measure of distance (cost function) between the logistic hypothesis function
\begin{equation}
\hbox{LR}_1(\mathbf{x})\equiv h_1^{(1)}(\mathbf{x})=\frac{1}{1+\exp[-f(\mathbf{x})]}
\end{equation}
and the true class labels $y$, where $y=0(1)$ if the event is a background(signal).
%, give by
%\begin{equation}
%\hbox{Cost}(f;y)=-\frac{1}{N}\sum_{I=1}^N \left[y_i\ln(h_1^{(1)}(\mathbf{x}))
%-h_1^{(1)}(\mathbf{x}_I)\right]+\lambda J(f)
%\end{equation}
%, $\lambda$ is a parameter controlling the regularization functional $J(f)$ in the residual sum-of-squares RSS~\cite{Hastie}, and the probability of classes are computed by the composition with the logistic hypothesis function
A regularization term helps to smooth out the output of the linear discriminant and reduce overfitting by penalizing too complex classification functions~\cite{Hastie, Duda}. Finally, by thresholding LR$_1$ we classify the events as signal if LR$_1>0.5$ and as background otherwise,  and compute the classification accuracies too. 
%We used the \sklearn~\cite{sklearn} implementation of the logistic regression classifier.

To train and test all the classifiers, we split 2 million BSM Higgs events of the data set~\cite{uciBaldi} provided by the authors of Ref.~\cite{Baldi:2014kfa} in four equal parts, 500000 events each. One of them was used to train and cross-validate all level-0 classifiers, another one to train and cross-validate the level-1 algorithm and getting the results of the level-0 algorithms, the third one for testing and getting the results of the generalizer. A fourth set was kept for future studies. This is just 20\% of the 11 million events used in training and testing their DNNs but we checked that with half this size, the results barely changed. Better results can be probably obtained by tuning the models parameters a bit more to compensate for a more restrictive regularization necessary to keep overfitting under control.

 The size of the data set was chosen in order that running the entire training, cross-validation and testing stages could be done rapidly after preparing the codes with a single 8-core Intel-i7 CPU. Apart from controlling overfitting by examining learning curves as the classification error in terms of the model complexity, we did not dispense too much time tuning the hyperparameters of our classifiers. 

 Overall, we took around 120 minutes to train, test and get the output scores of the three algorithms and of the generalizer. On the other hand, our implementation of a five hidden layers DNN with 300 units each with parameters tuned as in Ref.~\cite{Baldi:2014kfa} over 200 epochs with 10 million samples using \keras took 4680 minutes with our 8-core PC. Tuning a competitive DNN is far more difficult as, even with a tool like Bayesian optimization, no less than 100 runs is typically necessary to get hyperparameters close to the optimum choice. Of course, those running times would shrink considerably using more powerful resources. 
%By the way, the logistic regression algorithm has very few hyperparameters and we basically assumed the default settings from \sklearn. On the other hand, \xgboost and \keras  present many tuneable parameters that affect the classification performance of the algorithms. 

However, the cost to prepare a complex DNN is actually much larger than that demanded by stacking when we take into account the time to simulate the events themselves with an event generator like \texttt{MadGraph}~\cite{Alwall:2014hca}, for example. Simulating 11 million signal and background events with many final state particles as done in Ref.~\cite{Baldi:2014kfa} might take several days if the runs are not parallelized. In order to have good results, a DNN needs a large number of examples at the training stage. Our results, however, show that between 10 and 20\% of that number of examples is required by stacking only three algorithms to reach a performance not much worse than DNN and better than not stacked single algorithms. We discuss the results in details in the next section.

To train and test all the classifiers for the SM Higgs to tau leptons process, we split the entire data set~\cite{uciKaggle} provided by the ATLAS Collaboration in four equal parts, around 204500 events each. One of them was used to train and cross-validate all level-0 classifiers, another one to train and cross-validate the level-1 algorithm and getting the results of the level-0 algorithms, the third and fourth ones, totaling 409000 samples, for testing and getting the results of the level-1 generalizer. The size of these samples are very similar to those provided by ATLAS at the time of the Kaggle contest. In this case, tuning, training and testing \texttt{XGBoost} to reach the same performance of the HiggsML contest results lasted as much as the whole stacking procedure, once we did not need to use much time for tuning the algorithms for stacking.

We standardized the kinematic features used to train the level-0 classifiers by removing their medians and scaling the data according to the quantile range between the 1st quartile (25th quantile) and the 3rd quartile (75th quantile). This preprocessing was done with the \texttt{RobustScaler} routine of \sklearn.

It is important to stress that dispensing very careful tuning of hyperparameters is one of the advantages to be explored in stacking but it might possible to improve the overall performance by looking for the minimum classification error or the maximum AMS exploring the space of the parameters of the algorithms via a grid or random search, or even more efficiently using a Bayesian optimization algorithm. 
%Our experience shows that a very good performance can be achieved quickly and robustly by stacking compared to single algorithms which demand exhaustive parameter tunning.  

\subsection{Results} 
In the upper(lower) part of the Table~(\ref{tab1}), we show the algorithms used in the stacking procedure along with the classification accuracy, the area under curve (AUC) which measures the area below the Receiver-Operator Curve (ROC), the AMS evaluated at the cut which rejects(selects) 90(15)\% of backgrounds(signals), and the binned LLR statistics computed with $10^5$ pseudoexperiments for the BSM(SM) Higgs process. This selection criterion was chosen in order to match that of the Ref.~\cite{Baldi:2014kfa}(\cite{kaggle}).

Accuracy is measured as the ratio between the number of true signal identification plus the number of true background identification, and the total number of test examples. ROC is nothing but a curve of signal efficiency \emph{versus} background rejection. As the classification accuracy of the algorithm increases, AUC tends to one, a perfect classifier able to reject all backgrounds at the same time it retains all signal. 
\begin{table}[t] 
\centering
\begin{tabular}{ccccc} 
\hline\hline 
Classifier & Accuracy & AUC & AMS(cut) & LLR\\
\hline
 & & BSM $H\to H^\pm W^\mp$  & & \\ 
\hline
XGB$_0$ & 0.712 & 0.822 & {\bf 4.2(4.6)} & 1.8 \\
NN$_0$ & 0.717 & 0.792 & 3.7(4.1) & 4.1 \\ 
NB$_0$ & 0.601 & 0.662 & 2.2(2.4) & 2.3 \\ 
{\bf LR}$_1$  & {\bf 0.741} & {\bf 0.823} & {\bf 4.2(4.6)} & {\bf 5.5} \\
{\color{red}{\hbox{DNN}}} & & {\color{red}{0.885}} & {\color{red}{5.0(5.7)}} & \\
\hline
& & SM $h\to\tau^+\tau^-$  & & \\
\hline 
XGB$_0$ & 0.712 & 0.905 & 3.7 & 1.5 \\
NN$_0$ & 0.826 & 0.894 & 2.8 & 4.3 \\ 
NB$_0$ & 0.625 & 0.615 & 0.6 & 1.2 \\ 
{\bf LR}$_1$  & {\bf 0.844} & {\bf 0.912} & {\bf 4.0} & {\bf 5.4} \\
{\color{red}{\hbox{XGB}}} & & & {\color{red}{3.7}} & \\
\hline \hline                         
\end{tabular}
\caption{Classification accuracy, AUC, and signal significances for each level-0 and the level-1 ML classifiers computed with the $\hbox{AMS}_1$($\hbox{AMS}_2$) metrics of Eq.~(\ref{ams}) for BSM Higgs process and with $\hbox{AMS}_3$ of Eq.~(\ref{ams3}) for the SM Higgs process. The highest significances are marked in boldface. In red we show the significance of the DNN of Ref.~\cite{Baldi:2014kfa} and the XGB of Ref.~\cite{kaggle}. Accuracy is computed by classifying events with score $> 0.5$ as signals and those with score $< 0.5$ as background.} 
\label{tab1}
\end{table}
 Concerning the accuracy of classification, AUC and both AMS and binned LLR statistics, the generalizer performs better than BDT, the best level-0 classifier, in both processes despite the increase in significance for the the BSM Higgs process is marginal for the cut analysis as seen in Table~(\ref{tab1}). This is due the strong predominance of XGB for the performance of the stacked generalization in this case. It might be possible to improve the results by stacking more classifiers or replacing the weaker ones by better algorithms for this task. We found that 90\% background rejection is achievable at the cost of rejecting around 45(25)\% of signal events using the LR$_1$ output for the BSM(SM) Higgs search in the cut-and-count analysis. 
\begin{figure}[t]
\centering
\includegraphics[width=8.5cm]{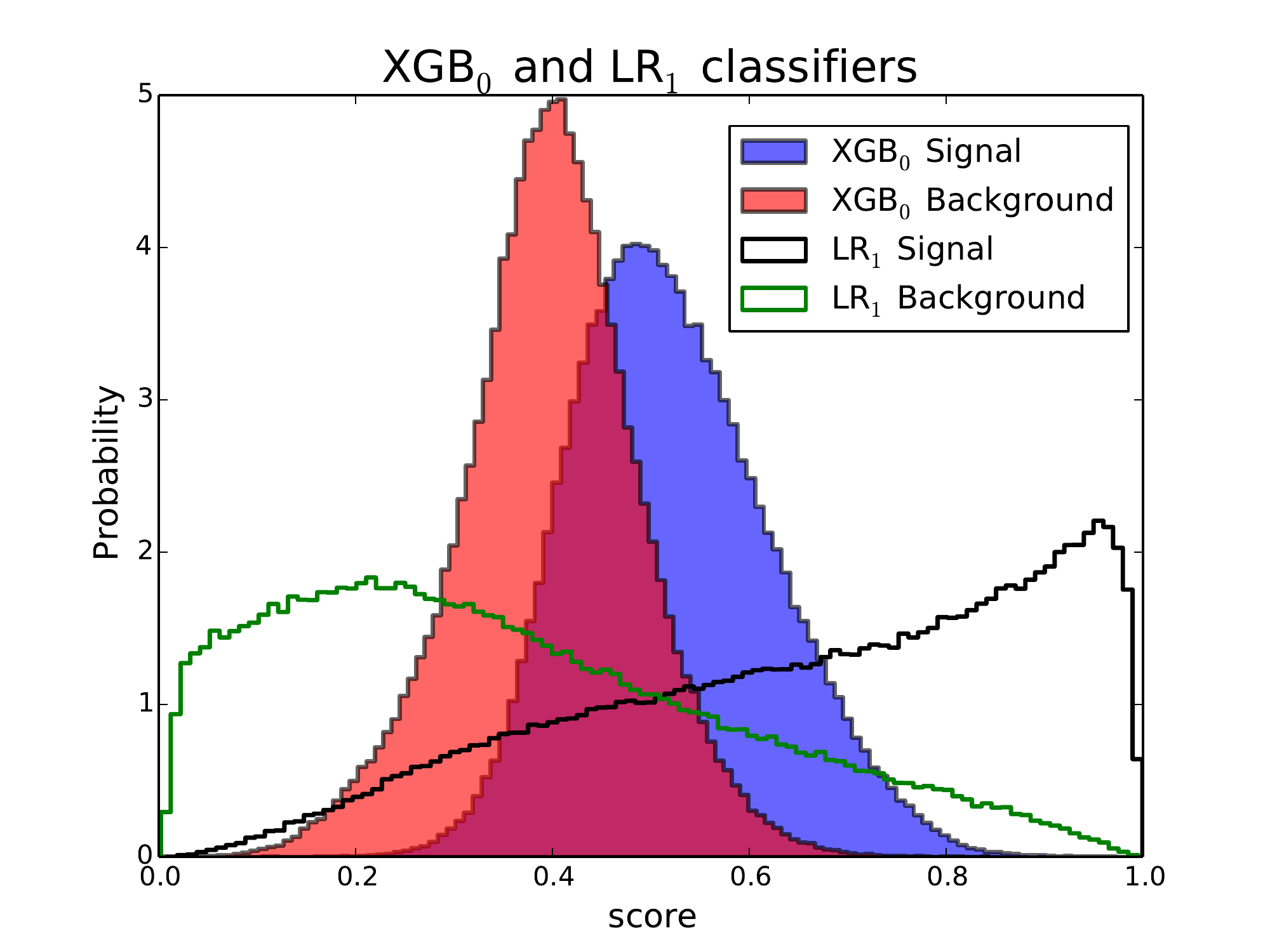}\\
\includegraphics[width=8.5cm]{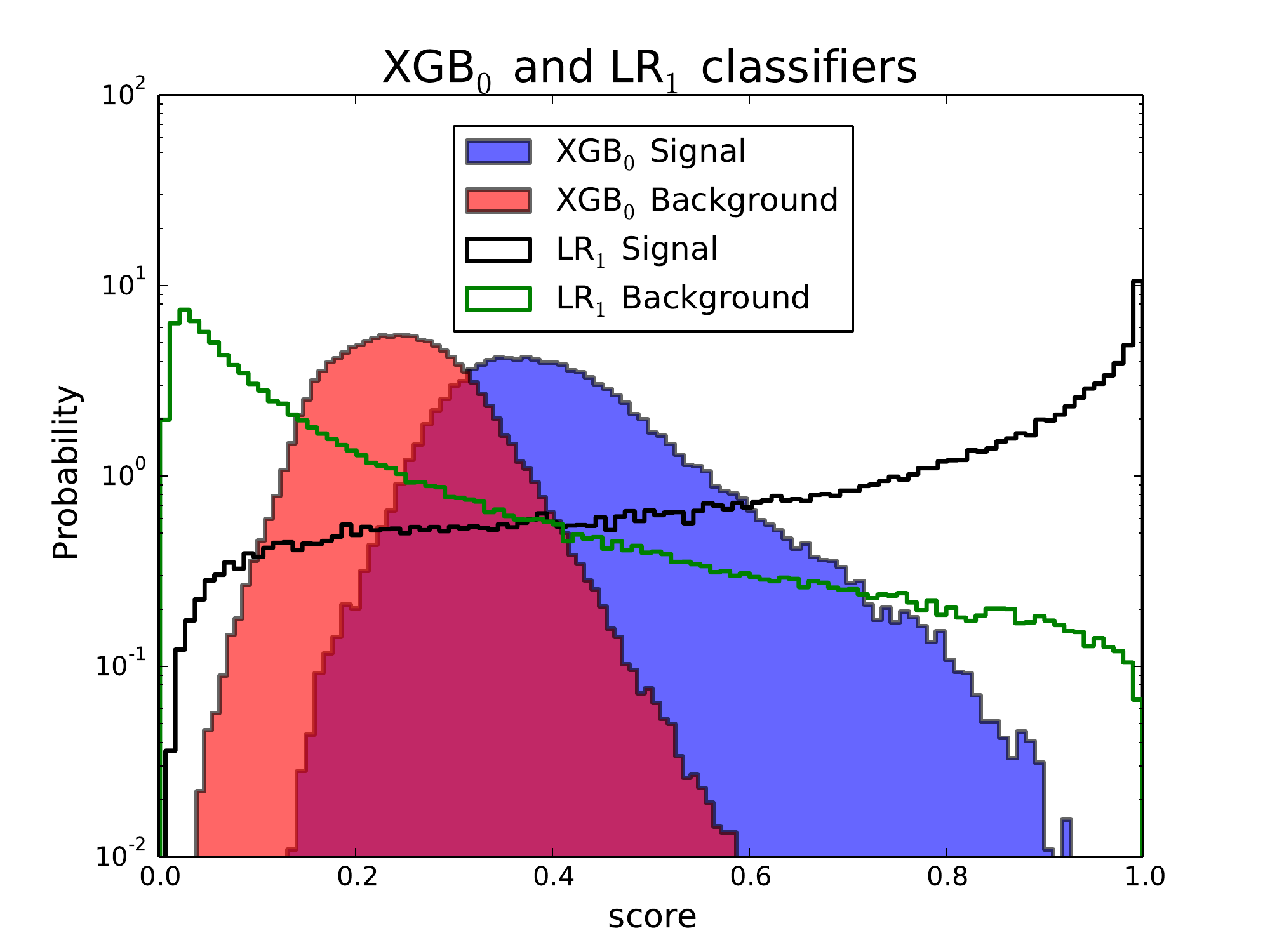}
\caption{Upper(lower) plot: the ML outputs distributions for the level-0 boosted decision trees XGB$_0$ from \xgboost, and the level-1 logistic regression classifier LR$_1$ from \sklearn for the BSM(SM) Higgs process.}
\label{fig:1}
\end{figure}

The best DNN of Ref.~\cite{Baldi:2014kfa} was able to reach $5(5.7)\sigma$ against $4.2(4.6)\sigma$ from LR$_1$ using the $\hbox{AMS}_1$($\hbox{AMS}_2$) metrics. For this calculation, we digitized the AUC curve presented in Ref.~\cite{Baldi:2014kfa} in order to obtain the signal efficiency and background rejection factors necessary to calculate the significance. It must be pointed out that stacking could be still be pushed much further than we did with little extra effort, probably stacking a larger number of simple classifiers enhances the performance as we said.

For the SM Higgs process, we reproduced the results of Ref.~\cite{chen}, reaching $\sim 3.7\sigma$ as shown in Table~(\ref{tab1}). The stacked generalizer, by its turn, was able to reach $4.2\sigma$ in the cut-and-count analysis, surpassing the results of Ref.~\cite{kaggle}.
%This is due our larger training set. The ATLAS Collaboration provided only around 250 thousand events for training at the time of the contest. After that, the collaboration released a larger dataset with more than 800 thousand events which we used in our work. 

%%% PARA AS CONCLUSÕES %%%%
%In spite of the fact that stacking is a very good ML tool to discriminate signals and backgrounds in this case, one might ask why not using a DNN to improve a bit more the classification performance and, ultimately, the statistical significances. The problem is that training, testing and optmizing a deep neural network currently demands powerful computational resources which might not be accessible to everyone as discussed in the previous section. In the lack of sufficient resources, stacking can be a very good alternative to train a highly discriminative ML tool.

In Fig.~(\ref{fig:1}) we show the outputs of the XGB$_0$ and LR$_1$ classifiers. The improvement in the discerning power of the generalizer compared to the best level-0 one, the BDT, is clear -- the output scores of background and signal events get much more concentrated towards 0 and 1, respectively. 

\begin{figure}[t]
\centering
\includegraphics[width=8.5cm]{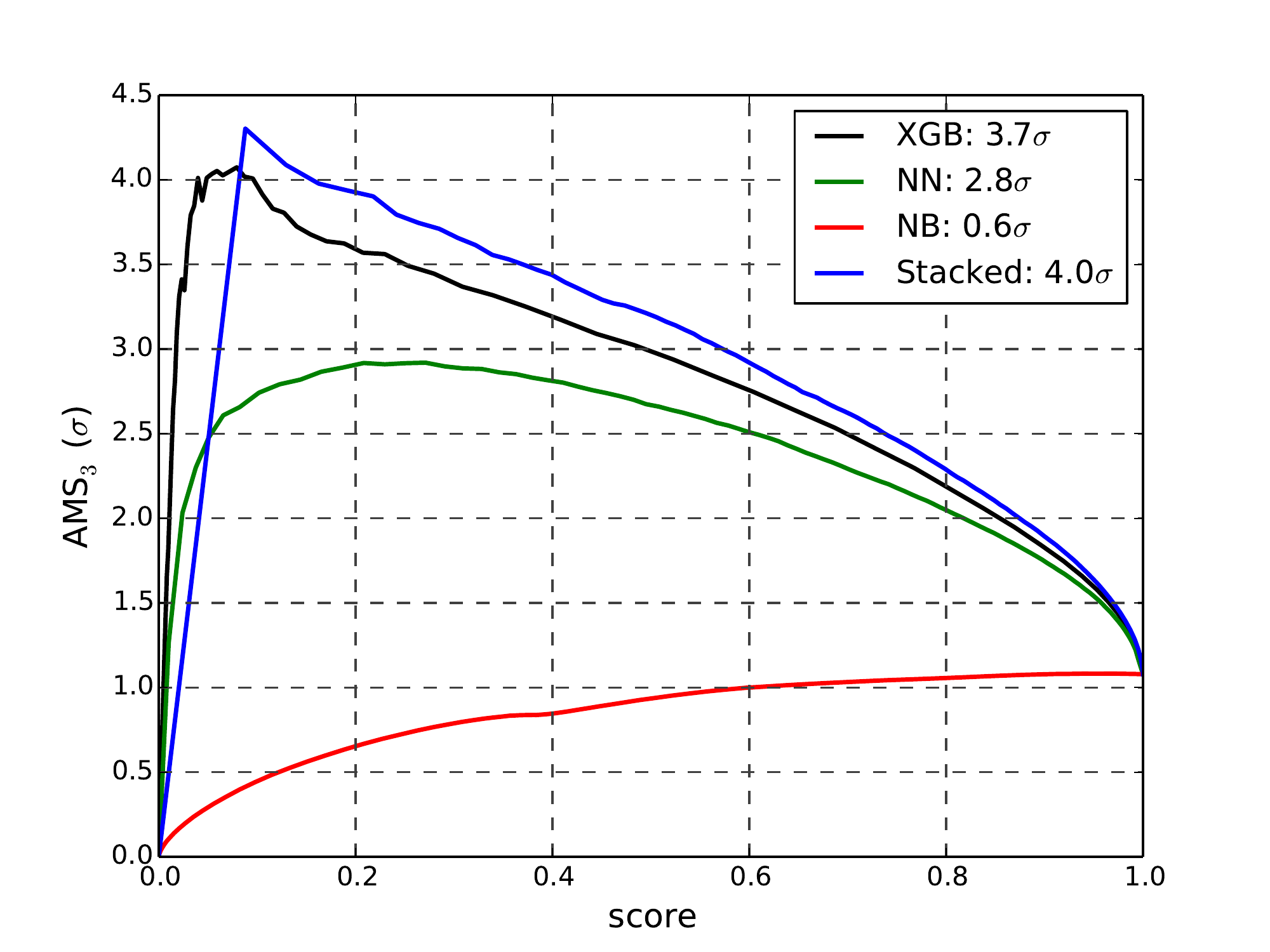}
\caption{The AMS$_3$ significances for three classifiers and the LR$_1$ generalizer obtained by stacking these classifiers in the SM Higgs study.}
\label{fig:ams}
\end{figure}

The gain of stacking the classifiers is better captured in Fig.~(\ref{fig:ams}) where we show $\hbox{AMS}_3$ in terms of the cut in the output score in SM Higgs process. First, the stacked LR$_1$ generalizer makes the curve more stable and less jagged than XGB, the second best classifier for that task. This instability of AMS in the cut analysis was reported by many teams of the HiggsML Kaggle contest. Second, stacking gives the biggest AMS$_3$, for all score cuts from 0.1 until 1. In special, for 0.15, the cut adopted in Ref.~\cite{chen}, it gives $4.0\sigma$ of significance.

The gain in significance is more evident in a multivariate analysis by estimating the binned log-likelihood ratio statistic distribution for both background and signal plus background hypothesis after marginalization over the systematic uncertainties (5\% in the number of background events). The boldface number of the last column of Table~(\ref{tab1}) displays the statistical significance of the LLR statistics based on the one dimensional LR$_1$ distribution. MVA improves significantly the results compared to the cut analysis. 
%The authors of Ref.~\cite{Baldi:2014kfa} did not perform a multivariate analysis thus a comparison is not possible in this case.
%It is not clear if stacking outperforms the DNNs of Ref.~\cite{Baldi:2014kfa} with MVA though.  

It might be difficult to estimate the systematic uncertainties on cross sections and the shape of the ML outputs but it is feasible as has been demonstrated in the single top experimental study~\cite{Aaltonen:2010jr,Likhomanenko:2015aba}. In our work we took just the uncertainty on the background normalization.

\section{Experiments with shallower neural networks}
In Refs.~\cite{Baldi:2014kfa,Baldi:2014pta}, it was shown that deep neural networks are not only very good discriminators but also less dependent on the features representation of the data set. Both in the SM Higgs search in the tau leptons channel~\cite{Baldi:2014pta} and the BSM Higgs decaying to $H^\pm W^\mp$~\cite{Baldi:2014kfa}, training DNNs with basic low-level distributions comprising just the raw data as delivered by the detectors as 4-momenta, or transverse momentum and rapidity, is almost as efficient as using high-level combinations of these basic features. In other words, DNNs learn the high-level man-made distributions to boost discoveries. 

The best DNN for $h\to\tau^+\tau^-$~\cite{Baldi:2014pta} used 7 hidden layers with 247 units each trained on 40 million examples, whereas for $H\to H^\pm W^\mp$~\cite{Baldi:2014kfa} the DNN consisted of 5 layers with 300 units each trained on more than 10 million instances. In both cases, a careful hyperparameters tuning was carried out. We emphasize that training, tuning and testing such complex DNNs can only be accomplished with larger computation facilities as clusters of CPUs or using GPUs. Unfortunately, we cannot compare our results for the SM Higgs process with those of Ref.~\cite{Baldi:2014pta} as the authors used somewhat different distributions, a much larger data set and different cross sections in their work but, summarizing, they found around $3.4\sigma$, calculated with a profile likelihood method, assuming 100($5000\pm 250$) signal(background) events.

%The performance comparison between DNNs trained on low and high-level features conducted in those works selected a particular set of high-level features which might not be the most powerful ones for discrimination. In this respect it is hard to say whether DNNs are really close to learn everything that can be learned in these classes of classification problems.

%As we said, stacking learners itself can be seen as a complex kind of feature engineering. 
Augmenting the features space of the learning problem with highly discriminating classifiers outputs might improve even further the discovery significance permitting to work with less complex DNN architectures.

In order to investigate this interesting possibility, we trained a 3 hidden layers neural network with 100 rectified linear units (ReLu) each  with~\keras to classify the events in both Higgs processes. We trained the DNNs over 100 epochs (with an early-stopping criterion) to keep overtraining under control after checking the classification error and log-loss curves as a function of iterations. We took 1 million(500 thousand) samples to train/validate(test) the classifier for the BSM study, and 204500(409) thousand for training/validation(testing) in the SM Higgs case. Hyperparameters of the DNNs were tuned 
using the Bayesian optimization routine \texttt{HyperOpt}~\cite{hyperopt} with 100 experiments. Running the 3-layer DNN for the BSM Higgs process took around  1008 minutes with an 8-core PC using the \texttt{Adam} optmizer.

We performed two experiments: (1) using the original feature spaces augmented with the level-0 output classifiers, (2) replacing all the kinematic features by the three level-0 ML outputs, that is, taking the DNN as the stacked generalizer. We show the comparisons in Table~(\ref{tab2}). 
%The 28(30) features used to train the level-0 algorithms in the $h\to\tau^+\tau^-$($H\to H^\pm W^\mp$) process passed to 31(33) after stacking the scores of the classifiers. 

%
\begin{table}[t] 
\centering
\begin{tabular}{ccccc} 
\hline\hline 
Classifier & Accuracy & AUC & AMS(cut) & LLR\\
\hline 
 & & BSM $H\to H^\pm W^\mp$  & & \\ 
\hline
LR$_1$  & 0.741 & 0.823 & 4.2(4.6) & 5.5 \\
DNN$_1$ & 0.740 & 0.822 & 4.0(4.4) & 7.4 \\
{\bf DNN } & {\bf 0.758} & {\bf 0.842} & {\bf 4.5(4.9)} & {\bf 8.1} \\
{\color{red}{\hbox{DNN}}} & & {\color{red}{0.885}} & {\color{red}{5.0(5.7)}} & \\
\hline
& & SM $h\to\tau^+\tau^-$   & & \\
\hline 
LR$_1$  & 0.843 & 0.914 & 4.2 & 5.4 \\
DNN$_1$ & 0.846 & 0.915 & 4.3 & 5.4 \\
{\bf DNN} & {\bf 0.848} & {\bf 0.917} & {\bf 4.5} & {\bf 5.6} \\
{\color{red}{\hbox{XGB}}} & & & {\color{red}{3.7}} & \\
\hline \hline                         
\end{tabular}
\caption{Comparison between the LR$_1$ generalizer and a 3 hidden layers neural network with 100 ReLu units each using the $\hbox{AMS}_1$($\hbox{AMS}_2$) metrics of Eq.~(\ref{ams}) for BSM Higgs process and with $\hbox{AMS}_3$ of Eq.~(\ref{ams3}) for the SM Higgs process. The highest significances are marked in boldface. In red we show the significance of the DNN of Ref.~\cite{Baldi:2014kfa} and the XGB of Ref.~\cite{kaggle}.}
\label{tab2}
\end{table}
Using only the level-0 outputs as features for our 3-layer DNN, that is, taking DNN$_1$ as the generalizer of the stacking, was not beneficial compared to the much simpler linear generalizer LR$_1$ in a cut analysis, however, it increased significantly the significance in MVA as we see in the second row of Table~(\ref{tab2}). In the third row we display the results of augmenting the original 28 features space with the three ML outputs used for stacking. In this case, the significance improves 7\% compared to LR$_1$ and it is only around 16\% worse than the large DNN of Ref.~\cite{Baldi:2014kfa} but, as antecipated, with much less computation efforts. Again, in MVA, this 3-layer DNN increased a lot the signal significance compared to all the other results, especially those of cut-and-count. We could not find any MVA analysis in Ref.~\cite{Baldi:2014kfa} to compare our results with but the lesson learned here is that allying less complex ML models with MVA is better than a complex model used for cut-and-count.
\begin{figure}[t]
\centering
\includegraphics[width=8.5cm]{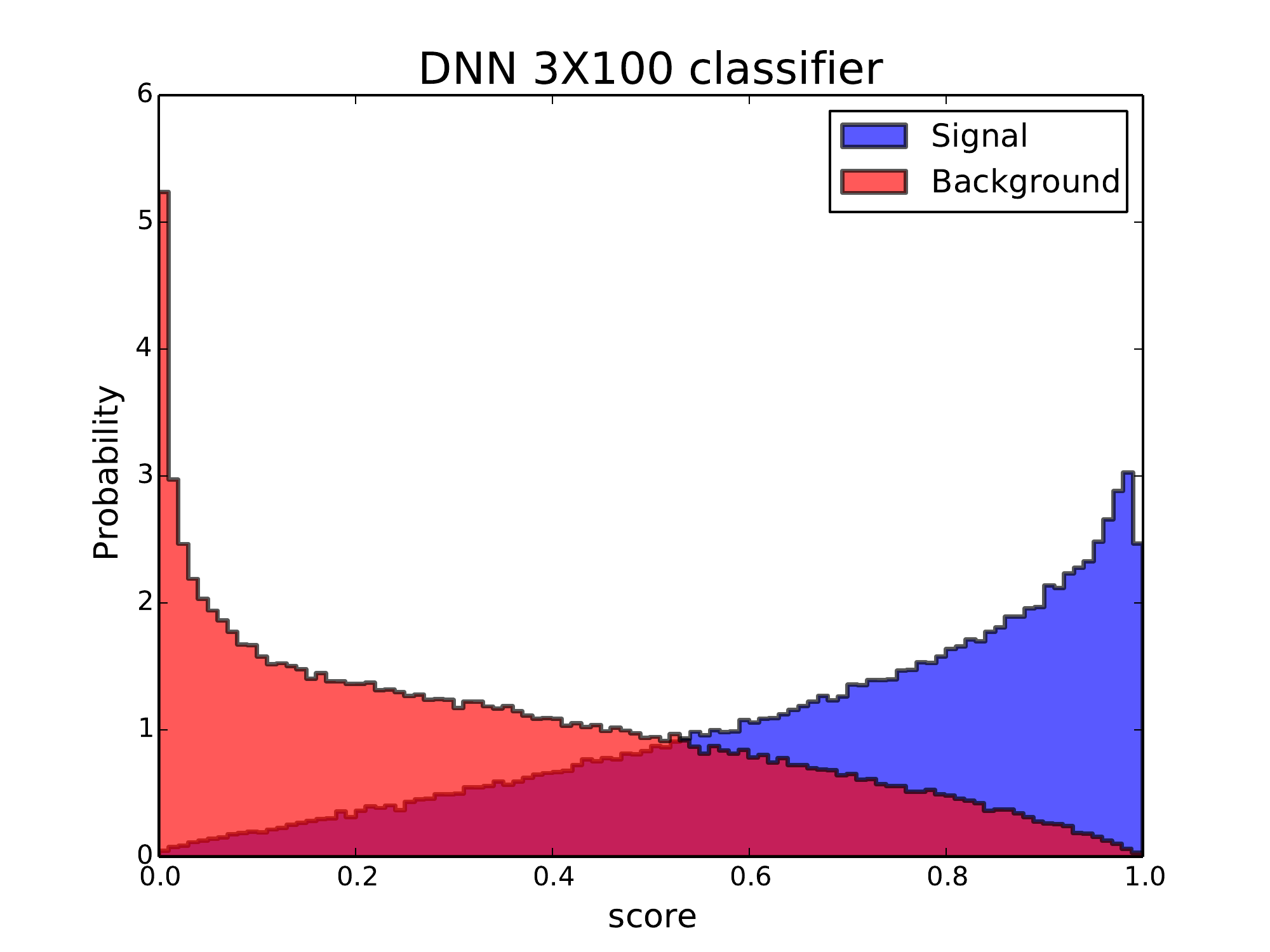}\\
\includegraphics[width=8.5cm]{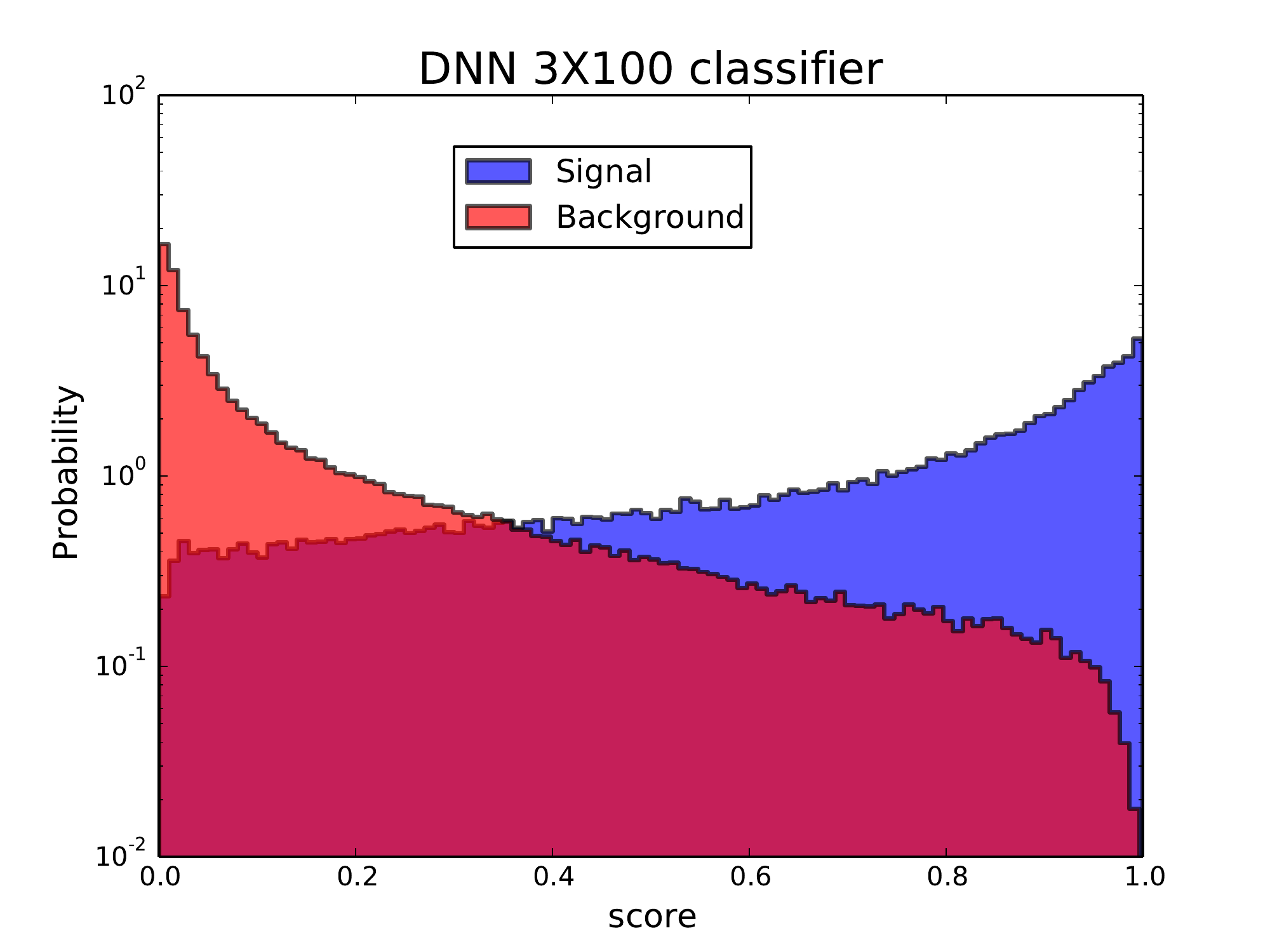}
\caption{Output scores for signals and backgrounds from a 3-hidden layers neural network with 100 units each built with \keras~\cite{keras} trained on a feature space augmented with the ML outputs of the stacking. In the upper(lower) plot we display the BSM(SM) Higgs process DNN output distribution.}
\label{fig:2}
\end{figure}
The new features used for stacking pays off the effort by demanding much less computation power and number of examples for training a good discriminant DNN. From our findings its fair to speculate that using more ML features and deeper neural networks might lead to even more impressive discrimination power reducing further the amount of data necessary for discoveries at high energy colliders.

Concerning the SM Higgs process, the 3-layer DNN presented better results for almost all approaches compared to LR$_1$ and the BDT of Ref.~\cite{chen} as can be checked in the Table~(\ref{tab2}).

In Fig.~(\ref{fig:2}) we show the output scores of our $3\times 100$ deep neural network for signal and backgrounds. It is clear that the distributions get even more squeezed towards the far ends compared of those of Fig.~(\ref{fig:1}) reflecting the better classification accuracy obtained with a DNN.

\section{Multivariate analysis with ML outputs}
What if instead we map the physical distributions to just one single ML output and perform the statistical analysis on it, we use all the level-0 and level-1 ML outputs in the MVA? As we stressed, it is likely that more heterogeneous algorithms are able to capture the correlations between the physical distributions necessary to improve the discernment between signal and backgrounds. This approach has been used by the LHCb Collaboration, for example, to efficiently tag $b$ and $c$-quark jets~\cite{Aaij:2015yqa}.

Using the ML outputs distributions to represent the data also brings another advantage. Sampling a low-dimensional PDF is much easier and faster than sampling a high-dimensional kinematic distribution PDF to calculate a binned likelihood ratio in MVA.
%Facilitating this task allows us to explore the correlations between the ML outputs which hopefully encode the correlations between the kinematic distributions used to build them. 
In Ref.~\cite{DeRujula:2010ys} the SM Higgs boson spin is discriminated among many hypothesis in the $ZZ\to 4\ell$ channel using the the angular distributions of the charged leptons. A binned LLR analysis was carried out with the fully correlated $N$-dimensional PDF ${\cal P}(x_1,\cdots,x_N)$ and the simplified PDF ignoring correlations given by $\prod_{i=1}^{N}{\cal P}(x_i)$, where $x_i$ denotes a kinematic distribution. The fully correlated analysis typically produces statistical significances twice as high compared to the approximated non-correlated PDF. 

One major difficulty when dealing with high-dimensional histograms is the binning itself. A high-dimensional histogram tends to be sparse with many empty bins which makes the computation hard. In this case, a more efficient binning procedure often reduces the dimensionality of the histogram losing information. In this respect, finding a good binning in higher dimensions might get easier, for example, by using an adaptive binning algorithm as proposed in ref.~\cite{binning}, where events in poorly populated regions are gathered in larger bins, and crowded regions can be more finely histogrammed, potentially augments the discernment power of the MVA.

%~\cite{tese}. 

\begin{figure}[t]
\centering
\includegraphics[width=8.5cm]{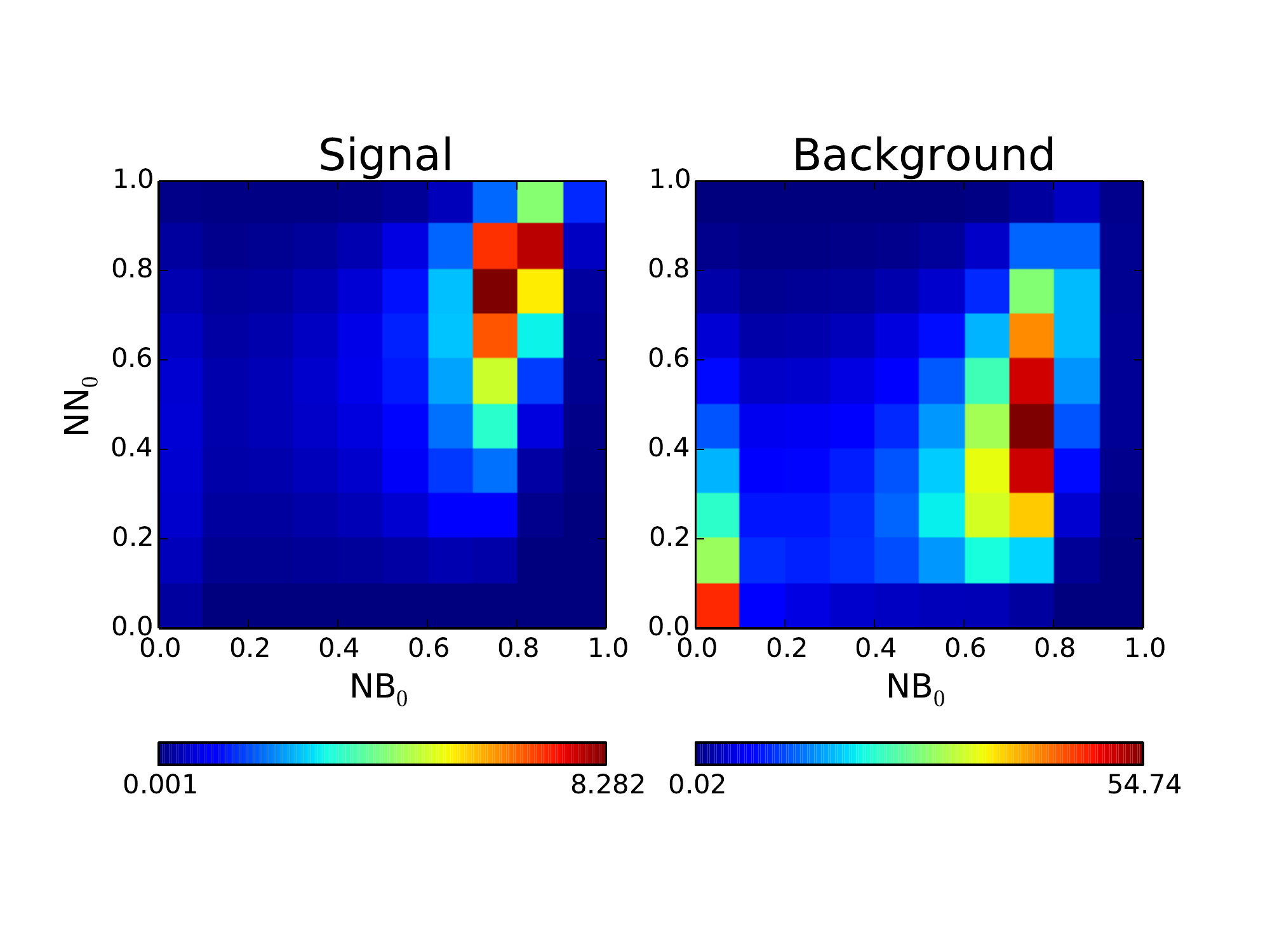} \\
\includegraphics[width=8.5cm]{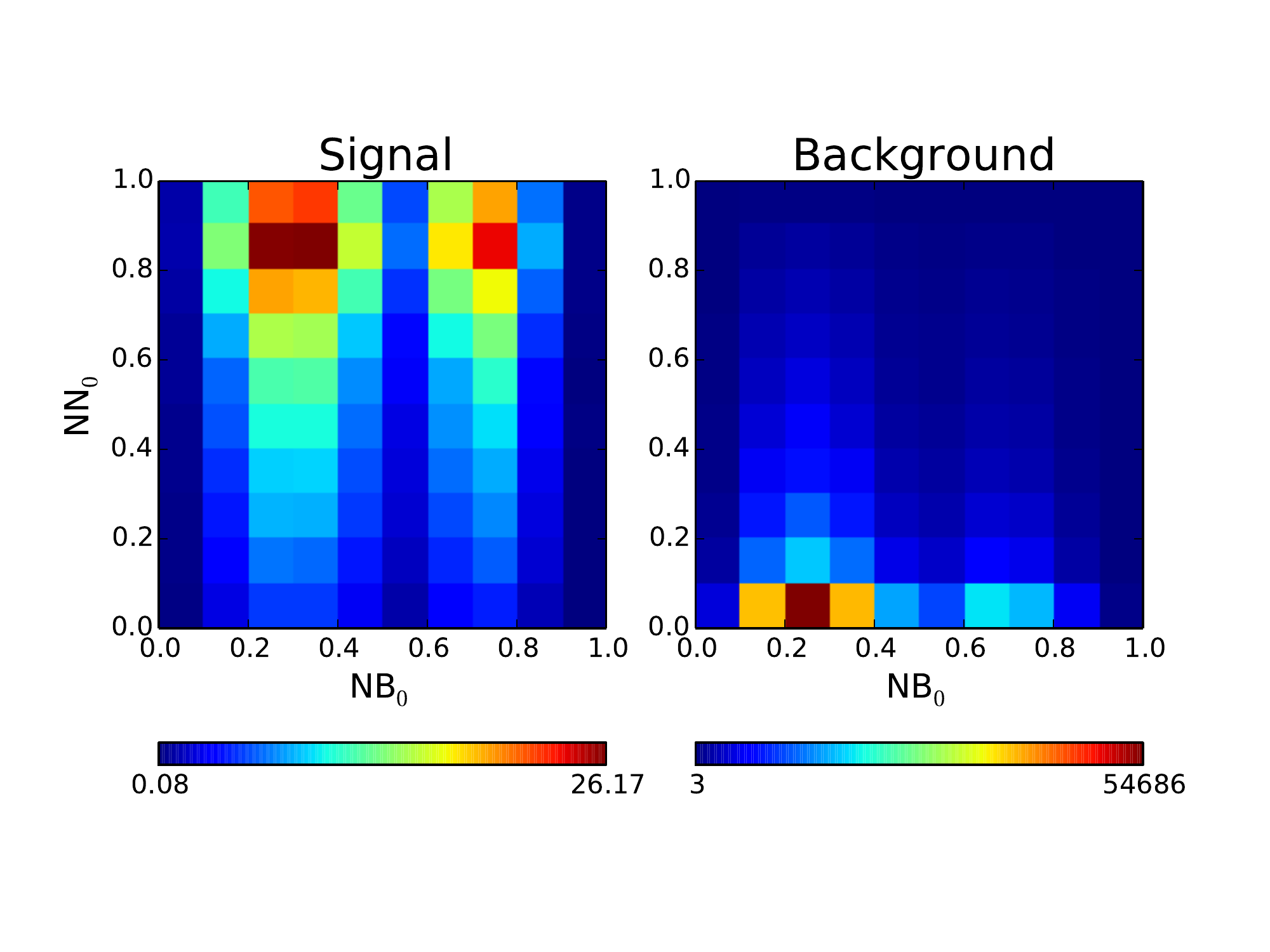}
\caption{The 2-dimensional color map of the NN$_0\times$NB$_0$ distribution ${\cal P}_{1\otimes 2}$ involved in computation of the results of Table~(\ref{tab3}). In the upper plot, the distribution of the $H\to H^\pm W^\mp$ events, and in the lower plot, the $h\to \tau^+\tau^-$ events.}
\label{fig:2d}
\end{figure}

With a 3-dimensional level-0 output space, we could explore, in principle, the full correlations between the ML outputs but, even this 3-dimensional PDF is already too sparse. Unless we adopt smarter binning methods, calculating LLR from the full PDF is difficult. However, we found that a $10\times 10$ histogram of the two least correlated outputs (in terms of the Pearson correlation), NN$_0$ and NB$_0$, is not sparse, so we chose to approximate the full PDF by ${\cal P}_{1\otimes 2}={\cal P}(h_1^{(0)})\times {\cal P}(h_2^{(0)},h_3^{(0)})$, that is, by ignoring the correlations between XGB$_0$ and the other two while maintaining the full correlations between NN$_0$ and NB$_0$. The results of the binned LLR calculation based on this PDF is presented in Table~(\ref{tab3}). In Fig.~(\ref{fig:2d}), we show the color map of ${\cal P}(\hbox{NN}_0,\hbox{NB}_0)$ for the BSM(SM) Higgs process in the upper(lower) plot, for signal and background in $10\times 10$ histograms.
\begin{table}[t] 
\centering
\begin{tabular}{c|c|c|c} 
\hline\hline 
Distributions & $h\to\tau^+\tau^-$ & $H\to H^\pm W^\mp$ & $H\to H^\pm W^\mp$(low) \\
\hline
only physical & 3.2 & 4.5 & 3.8 \\ 
level-0 & 4.7 & 4.8 & 3.6 \\
level-0+1 & 7.2 & 7.0 & 4.6\\ 
$\mathbf{1\otimes 2}$ {\bf PDF} & {\bf 8.0} & {\bf 9.5} & {\bf 6.0}\\
\hline \hline                         
\end{tabular}
\caption{Signal significance calculated with a binned log-likelihood ratio statistics of the distributions shown in the first column. The second column shows the results obtained with ML algorithms trained to discriminate SM Higgs to tau leptons events. The third(four) column displays LLR for the BSM Higgs process using algorithms trained with all(low-level) features.}
\label{tab3}
\end{table}

We computed the significance of the signal in MVA using: only the uncorrelated kinematic distributions (first row), only the uncorrelated level-0 outputs (second row), the uncorrelated level-0 plus level-1 ML output distribution, and the partially correlated ${\cal P}_{1\otimes 2}$ (last row of Table~(\ref{tab3})). As observed in Ref.~\cite{DeRujula:2010ys}, by exploring correlations between the features of the events large gains of about 2 in the significance is observed compared to the fully uncorrelated PDF (second row results) in both Higgs processes. Systematic uncertainties of 5\% in the backgrounds were taken into account in all computations. These same computations were carried out using ML outputs trained only on the low-level kinematic distributions of the BSM Higgs process (last column of Table~(\ref{tab3})). Remarkably, even in this case, discovery would possible by exploring the partially correlated PDF.

%The similarity between the results of the first and second rows of Table~(\ref{tab3}) suggests that the ML outputs can completely replace the kinematic distributions in MVA. This finding is in agreement with the results of~\cite{Cranmer:2015bka}. 
Note that augmenting the set of level-0 features with the level-1 is beneficial for the significance, maybe it is possible to continue the process of building ML outputs into large ensembles in order to use all the information about the process contained in ML outputs as is often done in many other data science problems.

\section{Conclusions}

%Machine learning applications have progressively shifted the paradigm of particle physics phenomenology into the artificial intelligence realm. 
State-of-art ML algorithms, as deep neural networks, have pushed the limits of the LHC to discover new particles and interactions to a new standard. Nevertheless, many other techniques and methods from the machine learning field are still waiting to be applied in particle physics. One of these techniques is ensemble learning, where various ML algorithms are trained to attack the same classification problem.

%Ensemble methods are common practice in data science and ML applications. 
In this work, we showed that \emph{stacking} only three ML algorithms is competitive against boosted decision trees but performs a little bit worse than deep neural networks. However, stacking presents advantages in training and testing time, computer resources and tuning of hyperparameters that might compensate this loss in performance compared to DNN once, in the processes which we studied in this work, stacking performed very well actually. It should be pointed out that stacking is an adjustable method, stacking more and possibly better algorihtms may increase its performance.

We compared our results against the deep neural network application in particle physics of Ref.~\cite{Baldi:2014kfa}, and the SM Higgs decaying to tau leptons results from the 2014 \emph{HEP meets ML} Kaggle contest~\cite{kaggle}.

We found that \emph{stacking} performed about 20\% worse than the complex DNNs of Ref.~\cite{Baldi:2014kfa} in a cut-and-count analysis based on the classifiers outputs distribution but with much less computation effort. Our ensemble comprised only three ML algorithms but it is common practice in data science to build large ensembles with tens of ML algorithms. We also trained a not so deep neural network augmenting the original features set with the level-0 ML outputs of the stacking. In this case, again with much less effort than needed for deeper architectures, we got a result only 16\% worse than Ref.~\cite{Baldi:2014kfa} with cut-and-count. 
%This also shows that stacking is an efficient feature engineering method.

In a multivariate statistical analysis using the binned log-likelihood ratio statistics, the logistic regression generalizer was able to reach more than $5\sigma$, surpassing the more complex 5-layer DNNs of Ref.~\cite{Baldi:2014kfa} with the cut-and-count analysis, and a 3-layer DNN generalizer was able to deliver $\sim 8\sigma$ of significance for the BSM Higgs process when trained with an augmented feature space blending the original kinematic distributions and the three level-0 ML outputs. These results show that allying MVA and simpler algorithms might be better than just cutting on the output of a more discriminating algorithm and that, apart from possible issues with systematic uncertainties, MVA plus state-of-art ML techniques as deep neural networks is probably the best approach of all.

In the case of the SM Higgs to tau leptons channel, stacking showed a higher statistical significance in a cut-and-count analysis compared to a carefully tuned boosted decision tree~\cite{chen} at the same time it stabilized the computation of the significance. In MVA, as in the BSM Higgs case, stacking improved much the prospects of discovering this Higgs channel at the 8 TeV LHC with the data accumulated by the time of Run I. A faster 3-layer DNN trained with a features set augmented by level-0 output classifiers also increased significantly the significances in this case.

The best results were obtained by computing the binned LLR from a partially correlated probability distribution function of the events in terms of the 3-dimensional PDF of level-0 ML scores. Because MVA performs much better when the correlations are taken into account, we found that BSM(SM) Higgs events, as considered in Ref.~\cite{Baldi:2014kfa}(\cite{chen}), can be discerned from backgrounds with $9.5(8.0)\sigma$ against $5(3.7)\sigma$ of the DNN(BDT) of that work. The dimensionality reduction achieved with the ML outputs representation of the data was crucial to obtain this result. This best approach implies a 75\% reduction in integrated luminosity to get the same significance of the best DNN study for the BSM Higgs process.

Based on these results, we conclude that stacking ML algorithms combined with MVA is a potentially more powerful approach to reduce the amount of data necessary for discoveries than using state-of-art algorithms in cut analysis. This approach brings more benefits as reducing the time for training, testing and tuning the ML algorithms, compared to DNN mainly. Of course, it is natural to investigate now how stacking can be combined with deep neural networks to achieve even better results in classification problems of particle physics. It should be stressed that once the computational resources become more powerful and cheaper, training DNNs tend to get faster, easier and more advantageous compared to other techniques, including stacking weaker learners. However, stacking can be used to improve even further the power of deep neural networks as suggested in ref.~\cite{Baldi:2014kfa}, turning the study of ensemble techniques important now and in the future.

%In the SM Higgs process, stacking more ML algorithms on the top of \texttt{XGBoost} increased the signal significance both in cut-and-count analysis and MVA but, again, with less care with tunning parameters. We also constructed a DNN architecture with 3 hidden layers in this case with

We believe that this investigation brings out a broader message: particle physics phenomenology can still benefit much from the vast variety of ML techniques commonly used in data science and artificial intelligence field in general. It has became important to evaluate the costs of updating the experimental analysis to incorporate powerful ML techniques as they can drastically reduce the amount of data necessary for particle searches. 
%We also advocate a more throughout change of paradigm in particle physics phenomenology from rectangular cuts towards machine learning applications by joining forces with data scientists and artificial inteligence experts.

\section*{Acknowledgments}
We thank the financial support from the Brazilian agencies CNPq (process 307098/2014-1), and FAPESP (process 2013/22079-8).
Ankit Patel~\cite{ankit} and Kuver Sinha kindly read the manuscript and made valuable suggestions and comments.

\end{document}